\documentclass[aps,prl,twocolumn,superscriptaddress]{revtex4}

\usepackage{graphicx}
\usepackage{dcolumn}
\usepackage{bm}

\begin{document}


\title{First order phase transition in the frustrated triangular antiferromagnet
CsNiCl$_3$.}


\author{G. Quirion, X. Han, M.L. Plumer}
\affiliation{Department of Physics and Physical Oceanography,
Memorial University, St. John's, Newfoundland, Canada, A1B 3X7}

\author{M. Poirier}
\affiliation{Regroupement Qu\'eb\'ecois sur les Mat\'eriaux de Pointe, D\'epartement de Physique,
Universit\'e de Sherbrooke, Sherbrooke, Qu\'ebec, Canada, J1K 2R1}


\date{\today}

\begin{abstract}
By means of high-resolution ultrasonic velocity measurements, as a function of temperature and
magnetic field, the nature of the different low temperatures magnetic phase transitions observed
for the quasi-one-dimensional compound CsNiCl$_3$ is established.  Special attention has been
devoted to the field-induced 120$^\circ$ phase transition above the multicritical point in the
$H-T$ phase diagram where the elastic constant $C_{44}$ reveals a step-like variation and
hysteresis effects. These results represent the first experimental evidence that the 120$^\circ$
phase transition is weakly first order and contradict the popular notion of new universality
classes for chiral systems.
\end{abstract}


\maketitle

Despite increasing interest in recent years in geometrically frustrated systems \cite{Moessner06},
there still exists considerable controversy over the nature of the phase transition in the
prototypical magnetic system characterized by near-neighbor antiferromagnetic exchange interactions
on a stacked triangular lattice \cite{Villian77, Toulouse77, Collins97}. The prediction made twenty
years ago that helical degeneracy associated with the 120$^\circ$ magnetic order leads to new
Heisenberg and XY chiral universality classes \cite{Kawamura87} found support over the following
ten years or so from numerous renormalization-group studies, Monte-Carlo simulations and
experimental data \cite{Collins97, Kawamura87, Kawamura98, aa, Weber95}. Results in favor of this
scenario continue to appear \cite{Plakhty00, DeFotis02, Peles03}. An alternative proposal of a very
weak fluctuation-induced first order transition made soon after the original suggestion
\cite{Azaria90} has also been strengthened by further theoretical studies and numerical simulations
\cite{Plumer97, Loison98, bb, Delamotte04, Peles04, Kanki06}. Experimental data on both rare-earth
helimagnets as well as ABX$_3$ compounds, such as CsMnBr$_3$, have been used extensively to support
each scenario. Evidence for a weak first-order transition was found in the thermal expansion data
on the helimagnet Ho long before this controversy surfaced \cite{Tindall77}. The results presented
in this Letter reveal the first experimental evidence that the 120$^\circ$ XY transition is weakly
first order. This is achieved also through magnetoelastic coupling effects, via ultrasonic sound
velocity measurements in the field-induced 120$^\circ$ phase of CsNiCl$_3$.

CsNiCl$_3$ is one of the more widely investigated of the large class of quasi-one-dimensional
hexagonal ABX$_3$ materials with strong antiferromagnetic $c$-axis exchange \cite{Collins97}. Along
with its sister compounds CsMnI$_3$, CsNiBr$_3$ and RbNiBr$_3$, it has  weak $c$-axis anisotropy
giving rise to a Linear (L) Ising-like ordered state in zero field at $T_{N_1}=4.8~$K, with an
additional in-plane ordering at $T_{N_2}=4.4~$K, resulting in an elliptical (E) polarization of the
spin density ({\bf S}) at low temperatures \cite{Plumer88}. These phases are characterized by a
period-2 modulation along the $c$-axis and period-3 in the basal plane. A magnetic field applied
along the $c$-axis of only 2.3~T is sufficient to induce a spin-flop phase where {\bf S} now lies
in the basal plane and forms the familiar 120$^\circ$ spin structure of the frustrated triangular
antiferromagnet. These three phases meet at an unusual type of multicritical point
\cite{Kawamura90} (at $T_m = 4.53~$K, $H_m = 2.3~$T). The transition to the Ising-like state
involves a phase-factor degeneracy associated with the triangular geometry and is been predicted to
belong to the XY universality class.  The nature of even this transition has a history of
controversy, with the most recent addition being an analysis of neutron diffraction data on a
similar transition in CsCoBr$_3$ suggesting tricritical behavior \cite{Mao02}. The transition at
$T_{N_2}$ is generally accepted to also be of XY universality.  At the multicritical point, the
axial anisotropy is exactly canceled by the applied field and the system is effectively isotropic
(Heisenberg-like). At higher field strengths, there is XY symmetry along with chiral degeneracy.
Such systems provide a convenient platform for the investigation of a line of phase transitions (to
the paramagnetic state) which can be sampled by changing the field strength.  In contrast with
present data, previous experimental investigations of this transition boundary support the notion
of $n=2$ chiral universality \cite{Collins97, Beckmann93, Enderle97}.

Magnetoelastic coupling has been repeatedly demonstrated to be a useful mechanism to reveal the
nature of the magnetic ordering in CsNiCl$_3$ \cite{Almond75, Johnson79, Poirier90}. Landau theory
can be used to show how elastic constants scale with the various order parameters and also to yield
mean-field predictions of anomalies at the transition boundaries \cite{Quirion05}. In the present
work, results and analysis are presented of high-resolution measurements of various elastic
constants as a function of both temperature and magnetic field, with a focus on the
paramagnetic-to-120$^\circ$ spin phase boundary. Step-like discontinuities are found where Landau
theory predicts none. Attempts at curve fitting to extract a critical exponent $\beta$ show that
this leads to values which are field dependent. The strongest evidence that the paramagnetic to
spin-flop phase boundary is weakly first order is found in the hysteretic behavior of $C_{44}$ as a
function of temperature and field.


Contrary to previous investigations \cite{Almond75, Poirier90, Trudeau92}, where ultrasonic
techniques were mainly used in order to determine phase diagrams, the emphasis of the present work
is on the measurement of critical phenomena in CsNiCl$_3$. This was realized using a
high-resolution pulsed ultrasonic interferometer to measure the temperature and magnetic field
dependence of different acoustic modes propagating along or perpendicular to the hexagonal
$c$-axis. Measurements were carried out on a single crystal specimen of 8.9~mm in length along the
$c$-axis and approximately 2.5~~mm along the perpendicular directions. The acoustic modes at 30~MHz
were generated using longitudinal and transverse lithium niobate transducers.


The analysis of the temperature or field dependence of elastic properties provides a convenient way
to study magnetic critical behavior in many systems. For hexagonal CsNiCl$_3$, a Landau type
approach has been used to determine the relationship between the variations in the elastic
constants $C_{33}$ and $C_{11}$ and the various order parameters \cite{Quirion05}. This coupling
occurs due to magnetoelastic contributions to the free energy of the form $\sim g~ e_i~ S^2$, where
$e_i$ is an element of the strain tensor and $S$ is the order parameter. A similar approach can be
used to include coupling terms which account for the application of a magnetic field along the
$c$-axis. One result of this new model \cite{Quirion06} indicates that the relative variation
$\Delta C_{33}/C_{33}$ can be generalized as
\begin{equation}\label{eq:C33S}
\frac{\Delta C_{33}(T,H)}{C_{33}} = - \Delta + \gamma~ S(T,H)^2
\end{equation}
where  $\Delta$ and $\gamma$ are constants specific to the magnetic transition of interest, while
the temperature and field dependencies are directly associated with those of the order parameter
$S$. According to (\ref{eq:C33S}), $C_{33}$ is expected to show a discontinuity even in the case of
a continuous phase transition. Thus, results on $C_{33}$ cannot be used to discriminate between a
continuous and weakly first order transition. This type of behavior is to be expected whenever a
linear-quadratic coupling, between the strain and the order parameter, is allowed by symmetry
\cite{Quirion05}. In order to clearly determine the character of the transition of the 120$^\circ$
phase, other ultrasonic modes need to be used, in particularly those that depend exclusively on
quadratic-quadratic coupling terms ($e^2 S^2$).  The allowed coupling terms, compatible with
hexagonal symmetry (not included in our previous analysis \cite{Quirion05}) are simply
\begin{eqnarray}
F_c(S,e_i) & = & \frac{g_{s_4}}{2}(e_4^2+e_5^2) S_z^2  +  \frac{g_{s_6}}{2}e_6^2 S_z^2 \\
\nonumber & + & \frac{g_{\beta_4}}{2}(e_4^2+e_5^2) S_\perp^2 + \frac{g_{\beta_6}}{2}e_6^2
S_\perp^2
\end{eqnarray}
where the notation of Ref.~[28] has been used. As these coupling terms are quadratic in
strain, the variation in the elastic constants can be written as
\begin{eqnarray}
\frac{\Delta C_{44}}{C_{44}} = \frac{\Delta C_{55}}{C_{55}} & = &g_{s_4} S_z^2 +
g_{\beta_4} S_\perp^2 \label{eq:C44TQ}\\
\frac{\Delta C_{66}}{C_{66}} & = & g_{s_6} S_z^2+ g_{\beta_6} S_\perp^2~.
\label{eq:C66TQ}
\end{eqnarray}
In the case of an hexagonal structure, $C_{44}$ and $C_{66}$ can be obtained by measuring the
velocity of transverse waves propagating along and perpendicular to the $c$-axis, respectively.

\begin{figure}[b]
 \includegraphics{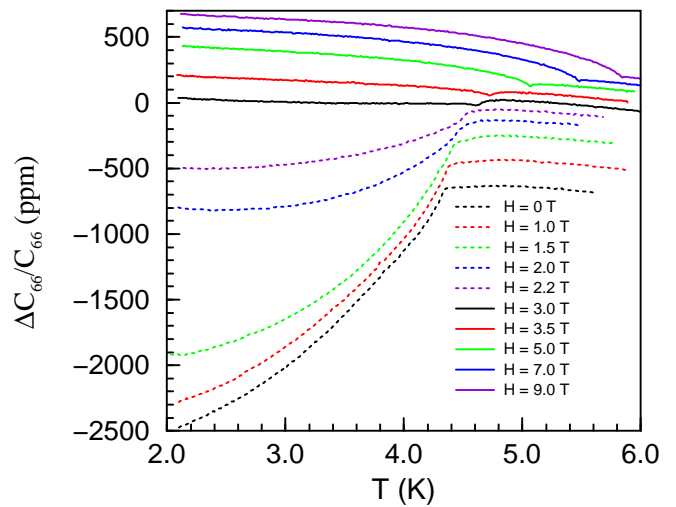}
    \caption{\label{C66} Relative variation of the elastic constant $C_{66}$
as a function of temperature. The broken and continuous lines represent results obtained below and
above $H_m = 2.3~$T, respectively.}
 \end{figure}

\begin{figure}[tb]
\includegraphics{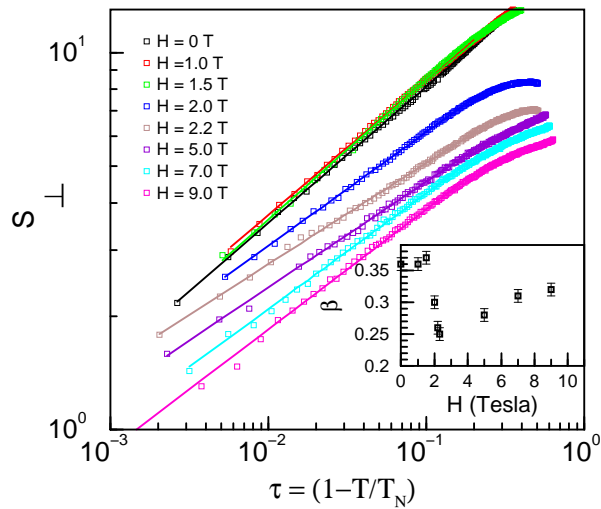}
\caption{\label{QC66} Order parameter as a function of the reduced temperature $\tau = 1-
T/T_{N_2}$ calculated using (\ref{eq:C66TQ}) and the results presented in Fig.~\ref{C66} . The data
obtained at different field are represented by symbols while lines represent fitted data at small
$\tau$ using a simple power law.  The inset shows the field dependence of the critical exponent
$\beta$ obtained from the power law fits.}
\end{figure}

Figure~\ref{C66} shows the most significant results of the temperature dependence of $\Delta
C_{66}/C_{66}$ with the magnetic field applied along the $c$-axis.
At $H = 0~$T, the onset of the L - E phase transition is clearly visible ($T_{N_2} = 4.33~$K) while
the variation at $T_{N_1}$ is barely noticeable.  The results show two distinct behaviors depending
on whether the value of the field is lower or higher than $H_m$. In the elliptical phase, $C_{66}$
softens as the temperature decreases while the opposite trend is observed in the 120$^\circ$ spin
phase above $H_m$. The observed temperature dependencies of $\Delta C_{66}/C_{66}$ are perfectly
consistent with the Landau predictions (\ref{eq:C66TQ}) and these data can be used to estimate the
temperature dependence of the order parameter $S_\bot$.  The results of this analysis, obtained at
different fields, are presented in Fig.~\ref{QC66} as a function of the reduced temperature $\tau =
1 - T/T_{N_2}$ on a log-log plot. All curves show a well defined power law behavior over a minimum
of two decades in $\tau$. Clearly, for $H < 1.5~$T a unique scaling is observed, confirming that
$\beta_E = 0.35 \pm 0.02$ is field independent in the elliptical phase. As the value of the field
approach $H_m$, the value of the critical exponent $\beta$ suddenly decreases and then gradually
increases at higher fields. The insert in Fig.~\ref{QC66} illustrates in detail how the value of
$\beta$ evolves as a function of a magnetic field for CsNiCl$_3$.
Very close to the multicrital point, $\beta$ reaches a minimum of $\beta = 0.25 \pm 0.02$ which
corresponds to the predicted value for $n=2$ chirallity \cite{Kawamura87} but is also close to that
of tricriticality, $\beta = 1/4$. At higher fields, $\beta$ increases significantly.
\begin{figure}[b]
 \includegraphics{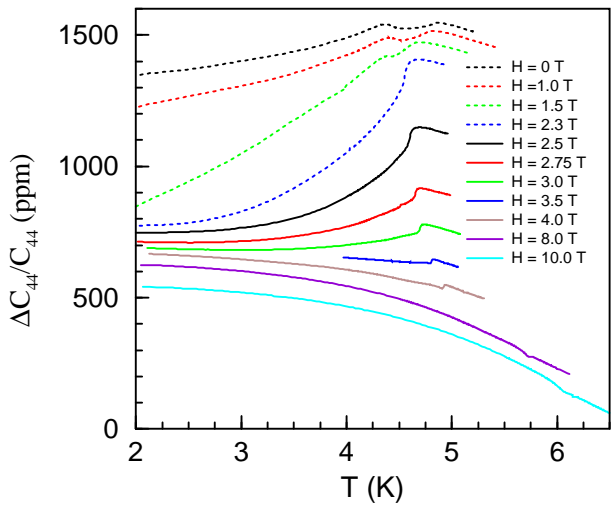}
    \caption{\label{C44} Relative variation of the elastic constant $C_{44}$
as a function of temperature. The broken and continuous lines represent results obtained below and
above $H_m = 2.3~$T, respectively.}
 \end{figure}
As noted previously \cite{Delamotte04, Peles04}, variation in the value of effective critical
exponents as a function of an irrelevant parameter (in this case, $H$) may indicate a weak
fluctuation-induced first order transition.

Close inspection of the upper curves presented in Fig.~\ref{C66} show an unexpected dip right at
the para-$120^\circ$ phase boundary. This very small anomaly persists at all fields above $H_m$ and
could be interpreted as the effect of a weakly first order transition. Motivated by these results,
additional measurements were made using transverse waves propagating along the $c$-axis and
polarized in the basal plane to obtain $C_{44}$. This series of results was obtained with an
exceptionally high resolution of 0.1~ppm. The data are presented in Fig.~\ref{C44} as a function of
temperature. At $H = 0$, two distinct features are noticeable and correspond to the onset of the
phase transitions at $T_{N_1}$ and $T_{N_2}$.  As the field is increased, both anomalies merge at
$H_m = 2.3~$T. Above the multicrital point (see continuous lines), the observed temperature
dependence changes within the 120$^\circ$ phase as the field increases. Moreover, no power law
relationship, as predicted by the Landau Model (\ref{eq:C44TQ}), could be identified. More
significantly, a step like variation is noticeable at the critical temperature. These observations
taken all together cannot be reconciliated with a continuous phase transition.

\begin{figure}[b]
\includegraphics{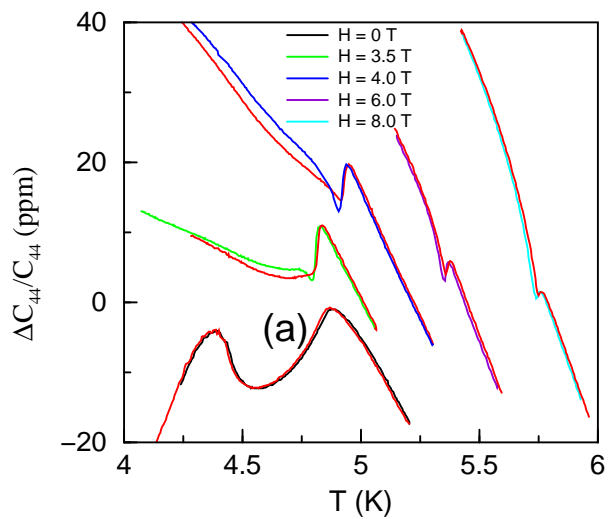}
\caption{\label{QC44} Thermal cycle analysis realized at different fields using the relative
variation of the elastic constant $C_{44}$. The data collected during the cooling process are all
represented in red color for clarity.  Curve (a) corresponds to the zero field transition at
$T_{N_2}$.  All other curves are associated with the transition to the 120$^\circ$ phase.}
\end{figure}
\begin{figure}[t]
 \includegraphics{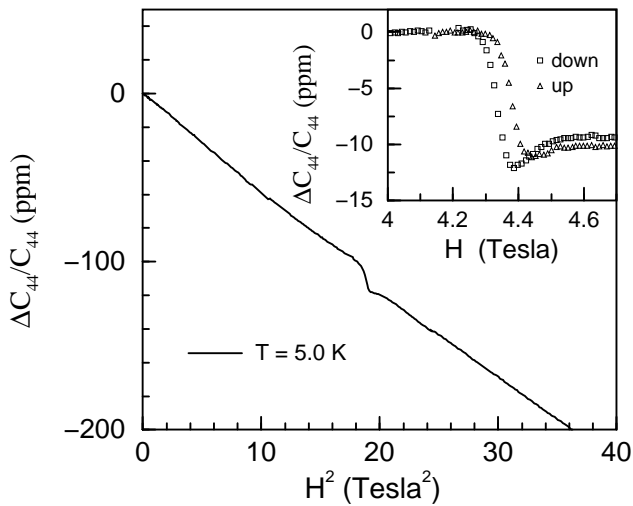}
    \caption{\label{C44H} Relative variation of the elastic constant $C_{44}$
as a function of $H^2$ measured at $T=5~$K. The inset shows the relative variation of $C_{44}$
after subtracting the magnetoelastic contribution observed in the paramagnetic phase.}
 \end{figure}

As a test of the first order character of the para-$120^\circ$ phase transition, the possibility of
thermal hysteresis was investigated. A collection of thermal cycles, realized using a
cooling/heating rate of 0.1~K/min, is presented in Fig.~\ref{QC44}.  The results obtained for the
$120^\circ$ phase  are compared to the data collected at $H = 0~$T. At zero field, where all
experimental evidence presented in this paper clearly indicate that the phase transition is
continuous, no significant hysteresis is observed. However, the transition to the $120^\circ$ spin
phase show a difference between the data collected during the heating and cooling processes. Those
differences are small but are systematically observed at all field values. The hysteresis is
maximum just below the critical temperature and persist over a temperature range of about 0.5~K. An
additional confirmation of first-order character is obtained from the field dependence of $C_{44}$.
The data collected at $T = 5.0~$K, presented in  Fig.~\ref{C44H} as a function of $H^2$, clearly
show the quadratic field dependence of $C_{44}$ in the paramagnetic phase.
This field dependence is associated with magnetostriction effects previously observed in CsNiCl$_3$
\cite{Rayne84}. The field dependence of $C_{44}$ near the 120$^\circ$ phase transition shown in the
insert of Fig.~\ref{C44H} has been isolated by subtracting this magnetostriction. It is clear that
the step like variation of $C_{44}$, along with the observed hysteresis, at the $120^\circ$ phase
boundary cannot be accounted for in the context of a continuous phase transition.

%



The data and analysis presented in this work serve to fill a long-standing gap regarding
experimental support for the growing body of theoretical and numerical work that the phase
transition in the prototypical geometrical frustrated system, the stacked triangular
antiferromagnet, is fluctuation-induced first order in nature.  CsNiCl$_3$ provides a convenient
field-temperature phase diagram for this purpose with a phase boundary line to the 120$^\circ$ spin
structure which may be sampled at various points. Sound velocity measurements have proven to be an
accurate tool for obtaining high resolution data on the various order parameters.  The possibility
to extract effective (field dependent) critical exponents, the weak nature of the discontinuities
and small hysteresis, taken together, provide evidence for the weakness of the first-order
character of this transition.  It is thus not surprising that conventional renormalization-group
techniques and Monte-Carlo simulations had previously been supportive of the notion of new chiral
universality classes associated with such frustrated systems.

We thank B. Southern for enlightening comments. This work was supported by grants from the Natural
Science and Engineering Research Council of Canada (NSERC) and Canada Foundation for Innovation
(CFI).

%





\end{document}